\documentclass[aps,prl,reprint,superscriptaddress,floatfix,longbibliography]{revtex4-1}

\usepackage{amsmath}    
\usepackage{amssymb}
\usepackage{graphicx}   
\usepackage[lofdepth,lotdepth, caption=false]{subfig}
\usepackage{verbatim}   
\usepackage{color}      
\usepackage{hyperref}   
\usepackage{dcolumn}
\usepackage{xcolor}  

\begin{document}
\newcommand{\psit}{(Pb$_{1-x}$Sn$_{x}$)$_{1-y}$In$_{y}$Te}
\newcommand{\pst}{Pb$_{1-x}$Sn$_{x}$Te}
\newcommand{\degrees}{$^\circ$C}

\title{Surface-state-dominated transport in crystals of the topological crystalline insulator In-doped Pb$_{1-x}$Sn$_x$Te}

\author{Ruidan~Zhong}
\email{rzhong@bnl.gov}
\affiliation{Condensed Matter Physics and Materials Science Department, Brookhaven National Laboratory, Upton, NY 11973, USA}
\affiliation{Materials Science and Engineering Department, Stony Brook University, Stony Brook, NY 11794, USA}
\author{Xugang~He}
\affiliation{Condensed Matter Physics and Materials Science Department, Brookhaven National Laboratory, Upton, NY 11973, USA}
\affiliation{Department of Physics and Astronomy, Stony Brook University, Stony Brook, NY 11794, USA}
\author{John~A.~Schneeloch}
\affiliation{Condensed Matter Physics and Materials Science Department, Brookhaven National Laboratory, Upton, NY 11973, USA}
\affiliation{Department of Physics and Astronomy, Stony Brook University, Stony Brook, NY 11794, USA}
\author{Cheng~Zhang}
\affiliation{Condensed Matter Physics and Materials Science Department, Brookhaven National Laboratory, Upton, NY 11973, USA}
\affiliation{Materials Science and Engineering Department, Stony Brook University, Stony Brook, NY 11794, USA}
\author{Tiansheng~Liu}
\affiliation{Condensed Matter Physics and Materials Science Department, Brookhaven National Laboratory, Upton, NY 11973, USA}
\affiliation{School of Chemical Engineering and Environment, North University of China, Shanxi 030051, China}
\author{Ivo~Pletikosi\'{c}}
\affiliation{Condensed Matter Physics and Materials Science Department, Brookhaven National Laboratory, Upton, NY 11973, USA}
\affiliation{Department of Physics, Princeton University, Princeton, NJ 08544, USA}
\author{Qiang~Li}
\author{Wei~Ku}
\author{Tonica~Valla}
\author{J.~M.~Tranquada}
\email{jtran@bnl.gov}
\author{Genda Gu}
\affiliation{Condensed Matter Physics and Materials Science Department, Brookhaven National Laboratory, Upton, NY 11973, USA}

\date{\today} 

\begin{abstract}
Three-dimensional topological insulators and topological crystalline insulators represent new quantum states of matter, which are predicted to have insulating bulk states and spin-momentum-locked gapless surface states.  Experimentally, it has proven difficult to achieve the high bulk resistivity that would allow surface states to dominate the transport properties over a substantial temperature range.  Here we report a series of indium-doped  Pb$_{1-x}$Sn$_x$Te compounds that manifest huge bulk resistivities together with strong evidence of topological surface states, based on thickness-dependent transport studies and magnetoresistance measurements. For these bulk-insulating materials, the surface states determine the resistivity for temperatures approaching 30 K.
\end{abstract}

\maketitle

A great deal of interest has been generated by the theoretical prediction and experimental realization of three-dimensional (3D) topological insulator (TI) materials \cite{Hasan2010,Qi2011}.  In certain semiconductors with strong spin-orbit coupling effects, the chiral character of metallic surface states is protected by time-reversal symmetry.  A variety of 3D TI materials have been synthesized over the last few years \cite{Weng2014}, and the existence of the topologically-protected surface states has been experimentally confirmed \cite{Hasan2010}; however, none of these materials have exhibited truly insulating bulk character. 

Topological crystalline insulators (TCIs) are the extension of (TIs) whose exotic surface states are protected by crystal symmetries, rather than by time-reversal symmetry \cite{Fu2011TCI}.   There has been considerable excitement since the first example, SnTe, was theoretically predicted \cite{Hsieh2012} and experimentally confirmed \cite{Tanaka2012} to exhibit topological surface states on \{001\}, \{110\} and \{111\} surfaces of the rock-salt crystal structure. Soon after this discovery, the topological surface states in the alloys Pb$_{1-x}$Sn$_{x}$Se and Pb$_{1-x}$Sn$_{x}$Te have been verified by angle-resolved photoemission spectroscopy (ARPES) \cite{Dziawa2012,Xu2012_Nature,Yan2014}, thus expanding the range of relevant materials.

For applications in spintronics, it is important to have the resistivity dominated by the topologically-protected surface states.   Substantial efforts have been made on the TI material Bi$_{2}$Se$_{3}$ and its alloys to reduce the bulk carrier density; however, while it has been possible to detect the signature of surface states in the magnetic-field dependence of the resistivity at low temperature \cite{Peng2009,Analytis2010,Xiong2012PRB}, attempts to compensate intrinsic defects \cite{Ren2011,Pan2014} have not been able to raise the bulk resistivity above 15~$\Omega$~cm.  Theoretical analysis suggests that even with perfect compensation of donor and acceptor defects, the resulting random Coulomb potential limits the achievable bulk resistivity \cite{Skinner2012}.

The solid solution Pb$_{1-x}$Sn$_{x}$Te provides a fresh opportunity for exploration.  The topological character changes from non-trivial at $x=1$ to trivial at $x=0$, with a topological quantum phase transition at $x_c\approx 0.35$, corresponding to the point at which band inversion onsets \cite{Tanaka2013, Dimmock1966, Gao2008}.  In our previous investigation of indium-induced superconductivity in Pb$_{0.5}$Sn$_{0.5}$Te single crystals \cite{Zhong2014}, we observed a non-monotonic variation in the normal-state resistivity with indium concentration, with a maximum at 6\%\ indium doping.  Further motivation has come from older work \cite{Shamshur2010} on various compositions of (Pb$_{1-x}$Sn$_x$)$_{1-y}$In$_y$Te.   Hence, we have performed a systematic study, growing and characterizing single crystals with six Pb/Sn ratios ($x= 0.2$, 0.25, 0.3, 0.35, 0.4, 0.5) and a variety of In concentrations ($y=0$--0.2).


\begin{figure*}
\centering
  \includegraphics[width=1.0\textwidth]{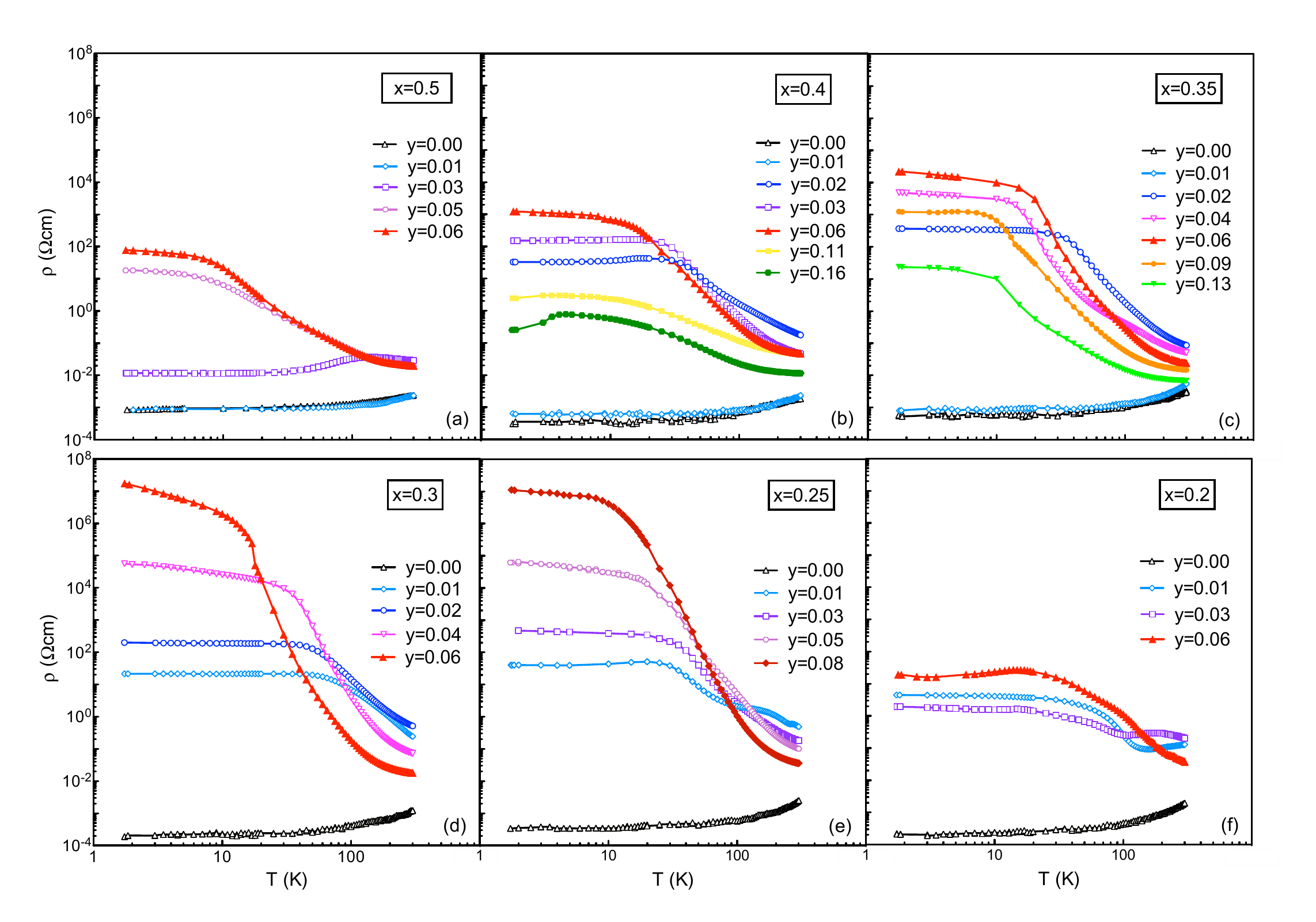}
   \caption{\label{fig:RT} (color online) Temperature dependence of resistivity in \psit\ for (a) $x=0.5$, (b) $x=0.4$, (c) $x=0.35$, (d) $x=0.3$, (e) $x=0.25$, and (f) $x=0.2$; the values of $y$ are labelled separately in each panel.  For each value of $x$, indium doping turns the metallic parent compound into an insulator, with low-temperature resistivity increasing by several orders of magnitude. The saturation of resistivity at temperatures below 30~K suggests that the surface conduction becomes dominant. }
\end{figure*}

Single crystal samples with nominal composition, \psit\ ($x_{\rm nom}= 0.2$--0.5, $y_{\rm nom}= 0$--0.2), were prepared via the modified Bridgeman method. Stoichiometric mixtures of high-purity elements [Pb (99.999\%), Sn (99.999\%), In (99.999\%), and Te (99.999\%)] were sealed in double-walled evacuated quartz ampoules. The ampoules were heated at 950\degrees\ in a box furnace and rocked to achieve good mixing of the ingredients. The crystal growth took place via slow cooling from 950 to 760\degrees\ in 1.5\degrees/hr, followed by gradual cooling to room temperature over another 3 days. Chemical composition values for $x$ and $y$ cited from here on correspond to the concentrations measured by energy-dispersive x-ray spectroscopy (EDS).
Nearly rectangular parallelepiped shaped samples were prepared by polishing, with a typical geometry of 5 mm long, 1.5 mm wide and 0.5 mm thick. Electrical resistance was measured in the standard four-probe configuration, using gold wires and room-temperature-cured, fast-drying silver paint for the ohmic contact on top side, performed with a Keithley digital multimeter (model 2001), where a Quantum Design Magnetic Property Measurement System was used for temperature control. Measurement errors due to the contact geometry are estimated to be less than 10\%. Sample thickness reduction is performed by sanding the bottom surface with the top contacts remaining nominally constant. Long-term relaxation in bulk resistance was observed in selected samples at low temperature, limited by the equilibration of carriers near In dopants \cite{Ravich2002}.  All magnetoresistance measurements were conducted after waiting for several half-lives (generally, 5 days for 5~K and 2 days for 20~K) so that the time-dependent component was negligible.

The measured resistivities, $\rho(T)$, for all samples, characterized by Sn concentration $x$ and In concentration $y$, are summarized in Fig.~\ref{fig:RT}. For each value of $x$, one can see that the resistivity of the parent compound ($y=0$, black open triangles) reveals weakly metallic behavior; furthermore, the magnitudes of $\rho$ in the In-free samples depend only modestly on $x$. With a minimum of $\sim$2\% indium doping, the low-temperature resistivity grows by several orders of magnitude, and the temperature dependence above $\sim30$~K exhibits the thermal activation of a semiconductor.  The saturation of the resistivity for $T\lesssim30$~K is consistent with a crossover to surface-dominated conduction. 


\begin{figure*}[t]
 \centering
    \includegraphics[width=2\columnwidth]{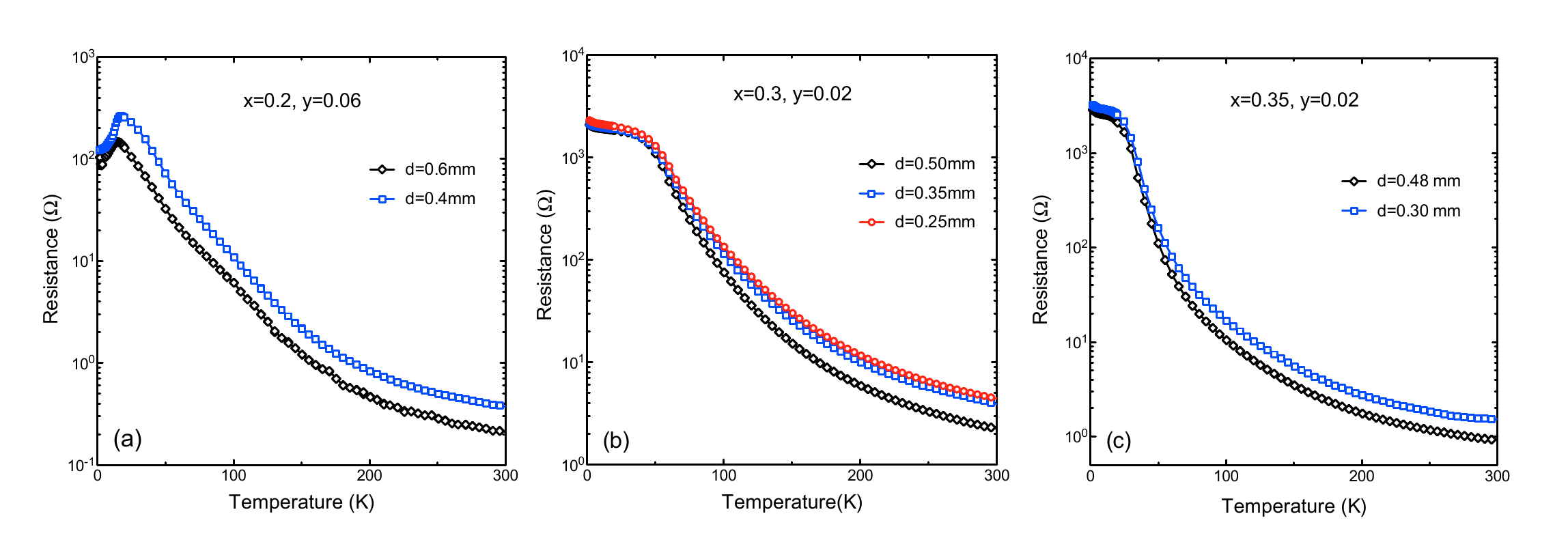}
    \caption{\label{fig:Thickness1} (color online)  Temperature dependent resistance of (a) (Pb$_{0.8}$Sn$_{0.2}$)$_{0.94}$In$_{0.06}$Te, (b) (Pb$_{0.7}$Sn$_{0.3}$)$_{0.98}$In$_{0.02}$Te, and (c)  (Pb$_{0.65}$Sn$_{0.35}$)$_{0.98}$In$_{0.02}$Te for varying sample thickness $d$. At high temperature, the resistance increases with decreasing thickness, while it becomes independent of $d$ at low temperature, consistent with conduction by surface states. }
\end{figure*}

The maximum resistivities, surpassing $10^6\ \Omega$ cm, are observed for $x=0.25$--0.3.  Even for $x=0.35$, doping with 6\%\ In results in a rise in resistivity of 6 orders of magnitude at low temperatures; higher In concentrations tend to result in a gradual decrease in $\rho$.   With increasing $y$, one eventually hits the solubility limit of In.  Exceeding that point results in an InTe impurity phase, which is superconducting below 4 K and appears to explain the low-temperature drop in resistivity for $x=0.4$ and $y=0.16$ illustrated in Fig.~\ref{fig:RT}(b).

Past studies \cite{Akimov1983, Ravich2002} of various transport properties in Pb$_{1-x}$Sn$_x$Te and the impact of In doping provide a basis for understanding the present results.  For In concentrations of $\lesssim0.06$, the In sites introduce localized states at a sharply defined energy that pins the chemical potential.  In a small range of Sn concentration centered about $x=0.25$, the chemical potential should be pinned within the band gap.  Hence, the very large bulk resistivities observed for $x=0.25$ and 0.3 are consistent with truly insulating bulk character.

Next, we test that the saturation of the resistivity at low temperature is due to surface-state conductivity.  
Following a recent study \cite{Kim2014} of topological surface states in SmB$_6$, we take advantage of the fact that the resistance due to surface transport should be independent of sample thickness, whereas the resistance from bulk transport should be inversely proportional to the thickness.  Figure~\ref{fig:Thickness1} shows measurements of resistance vs.\ temperature for various thicknesses of three different samples.  In each case, one can see that the resistance is essentially independent of thickness at low temperature, consistent with surface transport, whereas it increases with reduced thickness at higher temperatures, as expected for bulk transport.

\begin{figure}
 \centering
    \includegraphics[width=1.0\columnwidth]{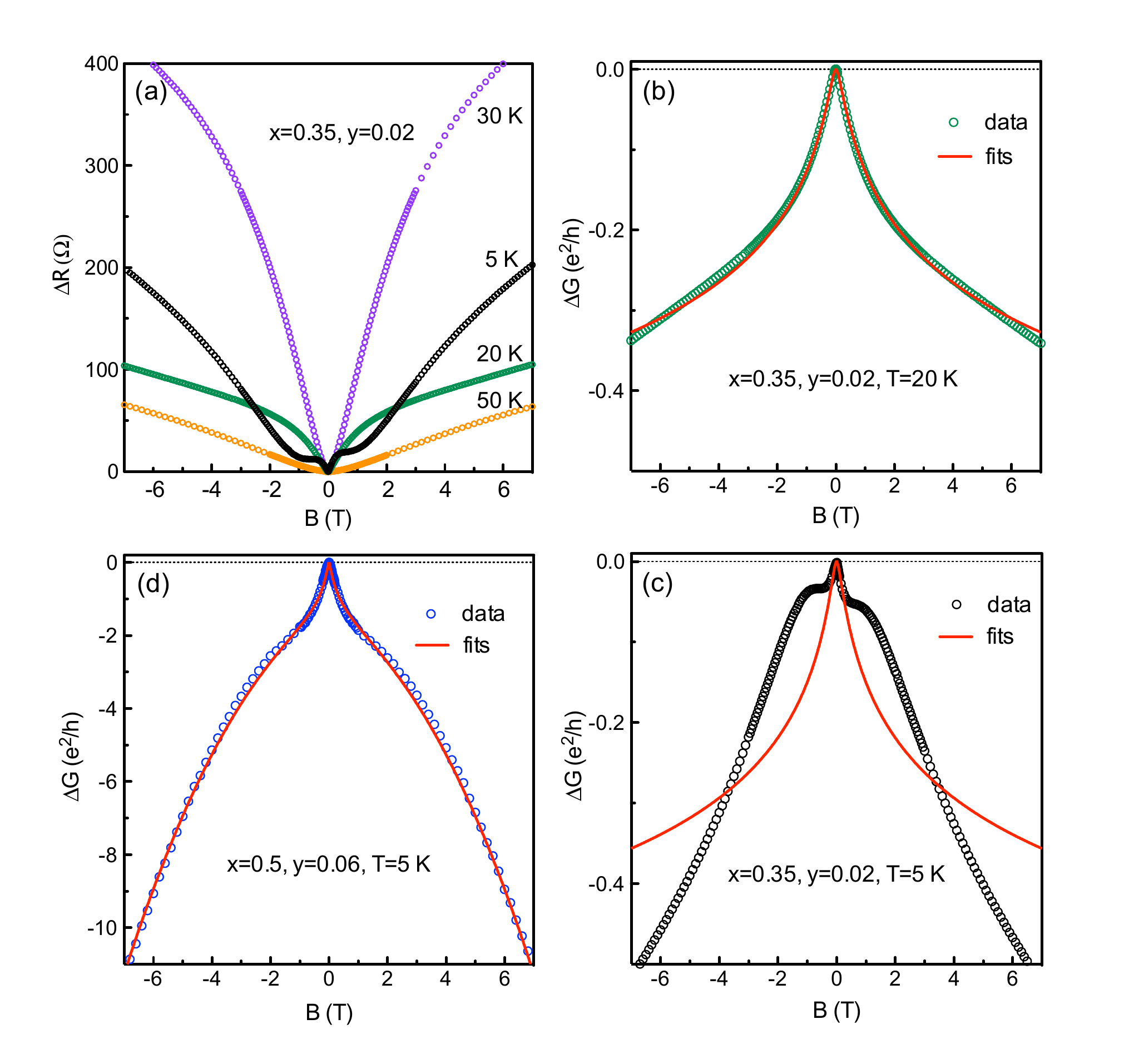}
    \caption{\label{fig:RH} (color online) (a) Magnetoresistance $\Delta R$ at $T= 5$, 20, 30 and 50 K for a (Pb$_{0.65}$Sn$_{0.35}$)$_{0.98}$In$_{0.02}$Te sample, with perpendicular magnetic field of $|B|\leq 7$~T. The WAL signature is observed at temperatures lower than 30 K. (b, c) Magnetoconductance, $\Delta G=\Delta (1/R)$, of the (Pb$_{0.65}$Sn$_{0.35}$)$_{0.98}$In$_{0.02}$Te sample measured at 20~K and 5~K, respectively. Lines represent fits to Eq.~(1), as discussed in the text. (d) Magnetoconductance of the (Pb$_{0.5}$Sn$_{0.5}$)$_{0.94}$In$_{0.06}$Te sample measured at 5~K. The line is a fit to Eq.~(1) together with an additional $-B^2$ term. }
\end{figure}

A common test of the topological character of surface states involves measurements of magnetoresistance (MR) at low temperature \cite{Bansal2012}.  The symmetry-protected coupling of spin and momentum for surface states makes them immune to weak localization effects.  Application of a transverse magnetic field violates the relevant symmetries \cite{Serbyn2014}, thus removing the topological protection and leading to a field-induced increase in resistance.   Figure~\ref{fig:RH}(a) shows data for $\Delta R=R(B)-R(0)$ measured at several temperatures for a magnetic induction of $|B|\leq 7$~T applied perpendicular to the surface of the (Pb$_{0.65}$Sn$_{0.35}$)$_{0.98}$In$_{0.02}$Te sample.  At temperatures of 30~K and below, the field dependence of the resistance has a form qualitatively consistent with that expected for weak anti-localization (WAL) of two-dimensional electron states.

To be more quantitative, we convert the data to conductance, $G$, and compare with the theoretical formula for WAL \cite{Hikami1980},
\begin{equation}
  \Delta G = {\alpha\over\pi} {e^2\over h} [\ln(B_\phi/B) - \psi({\textstyle\frac12}+B_\phi/B)],
\end{equation}
where $\psi$ is the digamma function and $\alpha$ is a number equal to $1/2$ times the number of conduction channels; $B_\phi= \Phi_0/(8\pi l_\phi^2)$, with $\Phi_0=h/e$ and $l_\phi$ being the electronic phase coherence length.   For our system, one expects 4 Dirac cones crossing the Fermi surface \cite{Liu2014,Assaf2014}, which would give $\alpha=2$.  Figure~\ref{fig:RH}(b) shows that we get a good fit to the 20-K data for the $x=0.35$, $y=0.02$ sample with $\alpha=0.37$ and $l_\phi=51$~nm.  Moving to $T=5$~K in Fig.~\ref{fig:RH}(c), the low-field data can be described by keeping $\alpha$ fixed and increasing $l_\phi$ to 58 nm; however, the data also exhibit a large oscillation about the calculated curve for $|B|>0.2$~T.  This may be due to a Landau level crossing the Fermi energy \cite{Serbyn2014}.  Turning to the $x=0.5$, $y=0.06$ sample, the 5-K data in Fig.~\ref{fig:RH}(d) are well described by the WAL formula for $|B|<1$~T, with $\alpha=2.25$ and $l_\phi=100$~nm, but at larger $|B|$ we need an additional component that varies as $-B^2$.  The latter contribution might come from bulk states.  In any case, for the samples tested, we have reasonable evidence for topological surface states.  We note that measurements on samples with $x=0.3<x_c$ indicate the absence of a robust WAL response.

We conclude that crystals of In-doped Pb$_{1-x}$Sn$_x$Te with a variety of compositions exhibit true bulk-insulating resistivity and metallic surface states that are topologically-protected for $x\gtrsim0.35$.   This allows one to exploit the unusual properties of the surface states in transport measurements without the need to apply a bias voltage to the surface.  Looking ahead, it is desirable to investigate the dispersion of the surface states with techniques such as angle-resolved photoemission\cite{Pletikosic2014}.   There is also strong interest in inducing topological superconductivity at surfaces or interfaces \cite{Qi2011}, and we note that there are exciting possibilities to create interfaces between the present topological insulators and superconductors of closely related alloys \cite{Zhong2014,Zhong2013}, such as In-doped SnTe and Pb$_{0.5}$Sn$_{0.5}$Te.

We thank W.-G. Yin for discussions. Work at Brookhaven is supported by the Office of Basic Energy Sciences, Division of Materials Sciences and Engineering, U.S. Department of Energy under Contract No.\ DE-SC00112704; use of facilities at the Center for Functional Nanomaterials is supported by the Office of Basic Energy Sciences, Division of Scientific User Facilities. I.P. is supported by the ARO MURI on Topological Insulators, grant W911NF-12-1-0461.


\begin{thebibliography}{30}%
\makeatletter
\providecommand \@ifxundefined [1]{%
 \@ifx{#1\undefined}
}%
\providecommand \@ifnum [1]{%
 \ifnum #1\expandafter \@firstoftwo
 \else \expandafter \@secondoftwo
 \fi
}%
\providecommand \@ifx [1]{%
 \ifx #1\expandafter \@firstoftwo
 \else \expandafter \@secondoftwo
 \fi
}%
\providecommand \natexlab [1]{#1}%
\providecommand \enquote  [1]{``#1''}%
\providecommand \bibnamefont  [1]{#1}%
\providecommand \bibfnamefont [1]{#1}%
\providecommand \citenamefont [1]{#1}%
\providecommand \href@noop [0]{\@secondoftwo}%
\providecommand \href [0]{\begingroup \@sanitize@url \@href}%
\providecommand \@href[1]{\@@startlink{#1}\@@href}%
\providecommand \@@href[1]{\endgroup#1\@@endlink}%
\providecommand \@sanitize@url [0]{\catcode `\\12\catcode `\$12\catcode
  `\&12\catcode `\#12\catcode `\^12\catcode `\_12\catcode `\%12\relax}%
\providecommand \@@startlink[1]{}%
\providecommand \@@endlink[0]{}%
\providecommand \url  [0]{\begingroup\@sanitize@url \@url }%
\providecommand \@url [1]{\endgroup\@href {#1}{\urlprefix }}%
\providecommand \urlprefix  [0]{URL }%
\providecommand \Eprint [0]{\href }%
\providecommand \doibase [0]{http://dx.doi.org/}%
\providecommand \selectlanguage [0]{\@gobble}%
\providecommand \bibinfo  [0]{\@secondoftwo}%
\providecommand \bibfield  [0]{\@secondoftwo}%
\providecommand \translation [1]{[#1]}%
\providecommand \BibitemOpen [0]{}%
\providecommand \bibitemStop [0]{}%
\providecommand \bibitemNoStop [0]{.\EOS\space}%
\providecommand \EOS [0]{\spacefactor3000\relax}%
\providecommand \BibitemShut  [1]{\csname bibitem#1\endcsname}%
\let\auto@bib@innerbib\@empty
\bibitem [{\citenamefont {Hasan}\ and\ \citenamefont {Kane}(2010)}]{Hasan2010}%
  \BibitemOpen
  \bibfield  {author} {\bibinfo {author} {\bibfnamefont {M.~Z.}\ \bibnamefont
  {Hasan}}\ and\ \bibinfo {author} {\bibfnamefont {C.~L.}\ \bibnamefont
  {Kane}},\ }\bibfield  {title} {\enquote {\bibinfo {title} {{Colloquium:
  Topological insulators}},}\ }\href@noop {} {\bibfield  {journal} {\bibinfo
  {journal} {Rev. Mod. Phys.}\ }\textbf {\bibinfo {volume} {82}},\ \bibinfo
  {pages} {3045--3067} (\bibinfo {year} {2010})}\BibitemShut {NoStop}%
\bibitem [{\citenamefont {Qi}\ and\ \citenamefont {Zhang}(2011)}]{Qi2011}%
  \BibitemOpen
  \bibfield  {author} {\bibinfo {author} {\bibfnamefont {Xiao-Liang}\
  \bibnamefont {Qi}}\ and\ \bibinfo {author} {\bibfnamefont {Shou-Cheng}\
  \bibnamefont {Zhang}},\ }\bibfield  {title} {\enquote {\bibinfo {title}
  {{Topological insulators and superconductors}},}\ }\href@noop {} {\bibfield
  {journal} {\bibinfo  {journal} {Rev. Mod. Phys.}\ }\textbf {\bibinfo {volume}
  {83}},\ \bibinfo {pages} {1057--1110} (\bibinfo {year} {2011})}\BibitemShut
  {NoStop}%
\bibitem [{\citenamefont {Weng}\ \emph {et~al.}(2014)\citenamefont {Weng},
  \citenamefont {Dai},\ and\ \citenamefont {Fang}}]{Weng2014}%
  \BibitemOpen
  \bibfield  {author} {\bibinfo {author} {\bibfnamefont {Hongming}\
  \bibnamefont {Weng}}, \bibinfo {author} {\bibfnamefont {Xi}~\bibnamefont
  {Dai}}, \ and\ \bibinfo {author} {\bibfnamefont {Zhong}\ \bibnamefont
  {Fang}},\ }\bibfield  {title} {\enquote {\bibinfo {title} {{Exploration and
  prediction of topological electronic materials based on first-principles
  calculations}},}\ }\href@noop {} {\bibfield  {journal} {\bibinfo  {journal}
  {MRS Bull.}\ }\textbf {\bibinfo {volume} {39}},\ \bibinfo {pages} {849--858}
  (\bibinfo {year} {2014})}\BibitemShut {NoStop}%
\bibitem [{\citenamefont {Fu}(2011)}]{Fu2011TCI}%
  \BibitemOpen
  \bibfield  {author} {\bibinfo {author} {\bibfnamefont {Liang}\ \bibnamefont
  {Fu}},\ }\bibfield  {title} {\enquote {\bibinfo {title} {{Topological
  Crystalline Insulators}},}\ }\href@noop {} {\bibfield  {journal} {\bibinfo
  {journal} {Phys. Rev. Lett.}\ }\textbf {\bibinfo {volume} {106}},\ \bibinfo
  {pages} {106802} (\bibinfo {year} {2011})}\BibitemShut {NoStop}%
\bibitem [{\citenamefont {Hsieh}\ \emph {et~al.}(2012)\citenamefont {Hsieh},
  \citenamefont {Lin}, \citenamefont {Liu}, \citenamefont {Duan}, \citenamefont
  {Bansil},\ and\ \citenamefont {Fu}}]{Hsieh2012}%
  \BibitemOpen
  \bibfield  {author} {\bibinfo {author} {\bibfnamefont {Timothy~H.}\
  \bibnamefont {Hsieh}}, \bibinfo {author} {\bibfnamefont {Hsin}\ \bibnamefont
  {Lin}}, \bibinfo {author} {\bibfnamefont {Junwei}\ \bibnamefont {Liu}},
  \bibinfo {author} {\bibfnamefont {Wenhui}\ \bibnamefont {Duan}}, \bibinfo
  {author} {\bibfnamefont {Arun}\ \bibnamefont {Bansil}}, \ and\ \bibinfo
  {author} {\bibfnamefont {Liang}\ \bibnamefont {Fu}},\ }\bibfield  {title}
  {\enquote {\bibinfo {title} {{Topological crystalline insulators in the SnTe
  material class}},}\ }\href@noop {} {\bibfield  {journal} {\bibinfo  {journal}
  {Nat. Commun.}\ }\textbf {\bibinfo {volume} {3}},\ \bibinfo {pages} {982}
  (\bibinfo {year} {2012})}\BibitemShut {NoStop}%
\bibitem [{\citenamefont {Tanaka}\ \emph {et~al.}(2012)\citenamefont {Tanaka},
  \citenamefont {Ren}, \citenamefont {Sato}, \citenamefont {Nakayama},
  \citenamefont {Souma}, \citenamefont {Takahashi}, \citenamefont {Segawa},\
  and\ \citenamefont {Ando}}]{Tanaka2012}%
  \BibitemOpen
  \bibfield  {author} {\bibinfo {author} {\bibfnamefont {Y.}~\bibnamefont
  {Tanaka}}, \bibinfo {author} {\bibfnamefont {Zhi}\ \bibnamefont {Ren}},
  \bibinfo {author} {\bibfnamefont {T.}~\bibnamefont {Sato}}, \bibinfo {author}
  {\bibfnamefont {K.}~\bibnamefont {Nakayama}}, \bibinfo {author}
  {\bibfnamefont {S.}~\bibnamefont {Souma}}, \bibinfo {author} {\bibfnamefont
  {T.}~\bibnamefont {Takahashi}}, \bibinfo {author} {\bibfnamefont {Kouji}\
  \bibnamefont {Segawa}}, \ and\ \bibinfo {author} {\bibfnamefont {Yoichi}\
  \bibnamefont {Ando}},\ }\bibfield  {title} {\enquote {\bibinfo {title}
  {{Experimental realization of a topological crystalline insulator in
  SnTe}},}\ }\href@noop {} {\bibfield  {journal} {\bibinfo  {journal} {Nat.
  Phys.}\ }\textbf {\bibinfo {volume} {8}},\ \bibinfo {pages} {800--803}
  (\bibinfo {year} {2012})}\BibitemShut {NoStop}%
\bibitem [{\citenamefont {Dziawa}\ \emph {et~al.}(2012)\citenamefont {Dziawa},
  \citenamefont {Kowalski}, \citenamefont {Dybko}, \citenamefont {Buczko},
  \citenamefont {Szczerbakow}, \citenamefont {Szot}, \citenamefont
  {{\L}usakowska}, \citenamefont {Balasubramanian}, \citenamefont {Wojek},
  \citenamefont {Berntsen}, \citenamefont {Tjernberg},\ and\ \citenamefont
  {Story}}]{Dziawa2012}%
  \BibitemOpen
  \bibfield  {author} {\bibinfo {author} {\bibfnamefont {P.}~\bibnamefont
  {Dziawa}}, \bibinfo {author} {\bibfnamefont {B.~J.}\ \bibnamefont
  {Kowalski}}, \bibinfo {author} {\bibfnamefont {K.}~\bibnamefont {Dybko}},
  \bibinfo {author} {\bibfnamefont {R.}~\bibnamefont {Buczko}}, \bibinfo
  {author} {\bibfnamefont {A.}~\bibnamefont {Szczerbakow}}, \bibinfo {author}
  {\bibfnamefont {M.}~\bibnamefont {Szot}}, \bibinfo {author} {\bibfnamefont
  {E.}~\bibnamefont {{\L}usakowska}}, \bibinfo {author} {\bibfnamefont
  {T.}~\bibnamefont {Balasubramanian}}, \bibinfo {author} {\bibfnamefont
  {B.~M.}\ \bibnamefont {Wojek}}, \bibinfo {author} {\bibfnamefont {M.~H.}\
  \bibnamefont {Berntsen}}, \bibinfo {author} {\bibfnamefont {O.}~\bibnamefont
  {Tjernberg}}, \ and\ \bibinfo {author} {\bibfnamefont {T.}~\bibnamefont
  {Story}},\ }\bibfield  {title} {\enquote {\bibinfo {title} {{Topological
  crystalline insulator states in Pb$_{1-x}$Sn$_x$Se}},}\ }\href@noop {}
  {\bibfield  {journal} {\bibinfo  {journal} {Nat. Mater.}\ }\textbf {\bibinfo
  {volume} {11}},\ \bibinfo {pages} {1023--1027} (\bibinfo {year}
  {2012})}\BibitemShut {NoStop}%
\bibitem [{\citenamefont {Xu}\ \emph {et~al.}(2012)\citenamefont {Xu},
  \citenamefont {Liu}, \citenamefont {Alidoust}, \citenamefont {Neupane},
  \citenamefont {Qian}, \citenamefont {Belopolski}, \citenamefont {Denlinger},
  \citenamefont {Wang}, \citenamefont {Lin}, \citenamefont {Wray},
  \citenamefont {Landolt}, \citenamefont {Slomski}, \citenamefont {Dil},
  \citenamefont {Marcinkova}, \citenamefont {Morosan}, \citenamefont {Gibson},
  \citenamefont {Sankar}, \citenamefont {Chou}, \citenamefont {Cava},
  \citenamefont {Bansil},\ and\ \citenamefont {Hasan}}]{Xu2012_Nature}%
  \BibitemOpen
  \bibfield  {author} {\bibinfo {author} {\bibfnamefont {Su-Yang}\ \bibnamefont
  {Xu}}, \bibinfo {author} {\bibfnamefont {Chang}\ \bibnamefont {Liu}},
  \bibinfo {author} {\bibfnamefont {N.}~\bibnamefont {Alidoust}}, \bibinfo
  {author} {\bibfnamefont {M.}~\bibnamefont {Neupane}}, \bibinfo {author}
  {\bibfnamefont {D.}~\bibnamefont {Qian}}, \bibinfo {author} {\bibfnamefont
  {I.}~\bibnamefont {Belopolski}}, \bibinfo {author} {\bibfnamefont {J.~D.}\
  \bibnamefont {Denlinger}}, \bibinfo {author} {\bibfnamefont {Y.~J.}\
  \bibnamefont {Wang}}, \bibinfo {author} {\bibfnamefont {H.}~\bibnamefont
  {Lin}}, \bibinfo {author} {\bibfnamefont {L.~A.}\ \bibnamefont {Wray}},
  \bibinfo {author} {\bibfnamefont {G.}~\bibnamefont {Landolt}}, \bibinfo
  {author} {\bibfnamefont {B.}~\bibnamefont {Slomski}}, \bibinfo {author}
  {\bibfnamefont {J.~H.}\ \bibnamefont {Dil}}, \bibinfo {author} {\bibfnamefont
  {A.}~\bibnamefont {Marcinkova}}, \bibinfo {author} {\bibfnamefont
  {E.}~\bibnamefont {Morosan}}, \bibinfo {author} {\bibfnamefont
  {Q.}~\bibnamefont {Gibson}}, \bibinfo {author} {\bibfnamefont
  {R.}~\bibnamefont {Sankar}}, \bibinfo {author} {\bibfnamefont {F.~C.}\
  \bibnamefont {Chou}}, \bibinfo {author} {\bibfnamefont {R.~J.}\ \bibnamefont
  {Cava}}, \bibinfo {author} {\bibfnamefont {A.}~\bibnamefont {Bansil}}, \ and\
  \bibinfo {author} {\bibfnamefont {M.~Z.}\ \bibnamefont {Hasan}},\ }\bibfield
  {title} {\enquote {\bibinfo {title} {{Observation of a topological
  crystalline insulator phase and topological phase transition in
  Pb$_{1-x}$Sn$_x$Te}},}\ }\href@noop {} {\bibfield  {journal} {\bibinfo
  {journal} {Nat. Commun.}\ }\textbf {\bibinfo {volume} {3}},\ \bibinfo {pages}
  {1192} (\bibinfo {year} {2012})}\BibitemShut {NoStop}%
\bibitem [{\citenamefont {Yan}\ \emph {et~al.}(2014)\citenamefont {Yan},
  \citenamefont {Liu}, \citenamefont {Zang}, \citenamefont {Wang},
  \citenamefont {Wang}, \citenamefont {Wang}, \citenamefont {Zhang},
  \citenamefont {Wang}, \citenamefont {Ma}, \citenamefont {Ji}, \citenamefont
  {He}, \citenamefont {Fu}, \citenamefont {Duan}, \citenamefont {Xue},\ and\
  \citenamefont {Chen}}]{Yan2014}%
  \BibitemOpen
  \bibfield  {author} {\bibinfo {author} {\bibfnamefont {Chenhui}\ \bibnamefont
  {Yan}}, \bibinfo {author} {\bibfnamefont {Junwei}\ \bibnamefont {Liu}},
  \bibinfo {author} {\bibfnamefont {Yunyi}\ \bibnamefont {Zang}}, \bibinfo
  {author} {\bibfnamefont {Jianfeng}\ \bibnamefont {Wang}}, \bibinfo {author}
  {\bibfnamefont {Zhenyu}\ \bibnamefont {Wang}}, \bibinfo {author}
  {\bibfnamefont {Peng}\ \bibnamefont {Wang}}, \bibinfo {author} {\bibfnamefont
  {Zhi-Dong}\ \bibnamefont {Zhang}}, \bibinfo {author} {\bibfnamefont {Lili}\
  \bibnamefont {Wang}}, \bibinfo {author} {\bibfnamefont {Xucun}\ \bibnamefont
  {Ma}}, \bibinfo {author} {\bibfnamefont {Shuaihua}\ \bibnamefont {Ji}},
  \bibinfo {author} {\bibfnamefont {Ke}~\bibnamefont {He}}, \bibinfo {author}
  {\bibfnamefont {Liang}\ \bibnamefont {Fu}}, \bibinfo {author} {\bibfnamefont
  {Wenhui}\ \bibnamefont {Duan}}, \bibinfo {author} {\bibfnamefont {Qi-Kun}\
  \bibnamefont {Xue}}, \ and\ \bibinfo {author} {\bibfnamefont
  {Xi}~\bibnamefont {Chen}},\ }\bibfield  {title} {\enquote {\bibinfo {title}
  {{Experimental Observation of Dirac-like Surface States and Topological Phase
  Transition in Pb$_{1-x}$Sn$_x$Te (111) Films}},}\ }\href@noop {} {\bibfield
  {journal} {\bibinfo  {journal} {Phys. Rev. Lett.}\ }\textbf {\bibinfo
  {volume} {112}},\ \bibinfo {pages} {186801} (\bibinfo {year}
  {2014})}\BibitemShut {NoStop}%
\bibitem [{\citenamefont {Peng}\ \emph {et~al.}(2010)\citenamefont {Peng},
  \citenamefont {Lai}, \citenamefont {Kong}, \citenamefont {Meister},
  \citenamefont {Chen}, \citenamefont {Qi}, \citenamefont {Zhang},
  \citenamefont {Shen},\ and\ \citenamefont {Cui}}]{Peng2009}%
  \BibitemOpen
  \bibfield  {author} {\bibinfo {author} {\bibfnamefont {Hailin}\ \bibnamefont
  {Peng}}, \bibinfo {author} {\bibfnamefont {Keji}\ \bibnamefont {Lai}},
  \bibinfo {author} {\bibfnamefont {Desheng}\ \bibnamefont {Kong}}, \bibinfo
  {author} {\bibfnamefont {Stefan}\ \bibnamefont {Meister}}, \bibinfo {author}
  {\bibfnamefont {Yulin}\ \bibnamefont {Chen}}, \bibinfo {author}
  {\bibfnamefont {Xiao-Liang}\ \bibnamefont {Qi}}, \bibinfo {author}
  {\bibfnamefont {Shou-Cheng}\ \bibnamefont {Zhang}}, \bibinfo {author}
  {\bibfnamefont {Zhi-Xun}\ \bibnamefont {Shen}}, \ and\ \bibinfo {author}
  {\bibfnamefont {Yi}~\bibnamefont {Cui}},\ }\bibfield  {title} {\enquote
  {\bibinfo {title} {{Aharonov-Bohm interference in topological insulator
  nanoribbons}},}\ }\href@noop {} {\bibfield  {journal} {\bibinfo  {journal}
  {Nat. Mater.}\ }\textbf {\bibinfo {volume} {9}},\ \bibinfo {pages} {225--229}
  (\bibinfo {year} {2010})}\BibitemShut {NoStop}%
\bibitem [{\citenamefont {Analytis}\ \emph {et~al.}(2010)\citenamefont
  {Analytis}, \citenamefont {McDonald}, \citenamefont {Riggs}, \citenamefont
  {Chu}, \citenamefont {Boebinger},\ and\ \citenamefont
  {Fisher}}]{Analytis2010}%
  \BibitemOpen
  \bibfield  {author} {\bibinfo {author} {\bibfnamefont {James~G.}\
  \bibnamefont {Analytis}}, \bibinfo {author} {\bibfnamefont {Ross~D.}\
  \bibnamefont {McDonald}}, \bibinfo {author} {\bibfnamefont {Scott~C.}\
  \bibnamefont {Riggs}}, \bibinfo {author} {\bibfnamefont {Jiun-Haw}\
  \bibnamefont {Chu}}, \bibinfo {author} {\bibfnamefont {G.~S.}\ \bibnamefont
  {Boebinger}}, \ and\ \bibinfo {author} {\bibfnamefont {Ian~R.}\ \bibnamefont
  {Fisher}},\ }\bibfield  {title} {\enquote {\bibinfo {title} {{Two-dimensional
  surface state in the quantum limit of a topological insulator}},}\
  }\href@noop {} {\bibfield  {journal} {\bibinfo  {journal} {Nat. Phys.}\
  }\textbf {\bibinfo {volume} {6}},\ \bibinfo {pages} {960--964} (\bibinfo
  {year} {2010})}\BibitemShut {NoStop}%
\bibitem [{\citenamefont {Xiong}\ \emph {et~al.}(2012)\citenamefont {Xiong},
  \citenamefont {Luo}, \citenamefont {Khoo}, \citenamefont {Jia}, \citenamefont
  {Cava},\ and\ \citenamefont {Ong}}]{Xiong2012PRB}%
  \BibitemOpen
  \bibfield  {author} {\bibinfo {author} {\bibfnamefont {Jun}\ \bibnamefont
  {Xiong}}, \bibinfo {author} {\bibfnamefont {Yongkang}\ \bibnamefont {Luo}},
  \bibinfo {author} {\bibfnamefont {YueHaw}\ \bibnamefont {Khoo}}, \bibinfo
  {author} {\bibfnamefont {Shuang}\ \bibnamefont {Jia}}, \bibinfo {author}
  {\bibfnamefont {R.~J.}\ \bibnamefont {Cava}}, \ and\ \bibinfo {author}
  {\bibfnamefont {N.~P.}\ \bibnamefont {Ong}},\ }\bibfield  {title} {\enquote
  {\bibinfo {title} {{High-field Shubnikov--de Haas oscillations in the
  topological insulator Bi$_2$Te$_2$Se}},}\ }\href@noop {} {\bibfield
  {journal} {\bibinfo  {journal} {Phys. Rev. B}\ }\textbf {\bibinfo {volume}
  {86}},\ \bibinfo {pages} {045314} (\bibinfo {year} {2012})}\BibitemShut
  {NoStop}%
\bibitem [{\citenamefont {Ren}\ \emph {et~al.}(2011)\citenamefont {Ren},
  \citenamefont {Taskin}, \citenamefont {Sasaki}, \citenamefont {Segawa},\ and\
  \citenamefont {Ando}}]{Ren2011}%
  \BibitemOpen
  \bibfield  {author} {\bibinfo {author} {\bibfnamefont {Zhi}\ \bibnamefont
  {Ren}}, \bibinfo {author} {\bibfnamefont {A.~A.}\ \bibnamefont {Taskin}},
  \bibinfo {author} {\bibfnamefont {Satoshi}\ \bibnamefont {Sasaki}}, \bibinfo
  {author} {\bibfnamefont {Kouji}\ \bibnamefont {Segawa}}, \ and\ \bibinfo
  {author} {\bibfnamefont {Yoichi}\ \bibnamefont {Ando}},\ }\bibfield  {title}
  {\enquote {\bibinfo {title} {{Optimizing Bi$_{2-x}$Sb$_x$Te$_{3-y}$Se$_y$
  solid solutions to approach the intrinsic topological insulator regime}},}\
  }\href@noop {} {\bibfield  {journal} {\bibinfo  {journal} {Phys. Rev. B}\
  }\textbf {\bibinfo {volume} {84}},\ \bibinfo {pages} {165311} (\bibinfo
  {year} {2011})}\BibitemShut {NoStop}%
\bibitem [{\citenamefont {Pan}\ \emph {et~al.}(2014)\citenamefont {Pan},
  \citenamefont {Wu}, \citenamefont {Angevaare}, \citenamefont {Luigjes},
  \citenamefont {Frantzeskakis}, \citenamefont {{de Jong}}, \citenamefont {{van
  Heumen}}, \citenamefont {Bay}, \citenamefont {Zwartsenberg}, \citenamefont
  {Huang}, \citenamefont {Snelder}, \citenamefont {Brinkman}, \citenamefont
  {Golden},\ and\ \citenamefont {{de Visser}}}]{Pan2014}%
  \BibitemOpen
  \bibfield  {author} {\bibinfo {author} {\bibfnamefont {Y.}~\bibnamefont
  {Pan}}, \bibinfo {author} {\bibfnamefont {D.}~\bibnamefont {Wu}}, \bibinfo
  {author} {\bibfnamefont {J.~R.}\ \bibnamefont {Angevaare}}, \bibinfo {author}
  {\bibfnamefont {H.}~\bibnamefont {Luigjes}}, \bibinfo {author} {\bibfnamefont
  {E.}~\bibnamefont {Frantzeskakis}}, \bibinfo {author} {\bibfnamefont
  {N.}~\bibnamefont {{de Jong}}}, \bibinfo {author} {\bibfnamefont
  {E.}~\bibnamefont {{van Heumen}}}, \bibinfo {author} {\bibfnamefont {T.~V.}\
  \bibnamefont {Bay}}, \bibinfo {author} {\bibfnamefont {B.}~\bibnamefont
  {Zwartsenberg}}, \bibinfo {author} {\bibfnamefont {Y.~K.}\ \bibnamefont
  {Huang}}, \bibinfo {author} {\bibfnamefont {M.}~\bibnamefont {Snelder}},
  \bibinfo {author} {\bibfnamefont {A.}~\bibnamefont {Brinkman}}, \bibinfo
  {author} {\bibfnamefont {M.~S.}\ \bibnamefont {Golden}}, \ and\ \bibinfo
  {author} {\bibfnamefont {A.}~\bibnamefont {{de Visser}}},\ }\bibfield
  {title} {\enquote {\bibinfo {title} {{Low carrier concentration crystals of
  the topological insulator Bi$_{2-x}$Sb$_x$Te$_{3-y}$Se$_y$: a
  magnetotransport study}},}\ }\href@noop {} {\bibfield  {journal} {\bibinfo
  {journal} {New J. Phys.}\ }\textbf {\bibinfo {volume} {16}},\ \bibinfo
  {pages} {123035} (\bibinfo {year} {2014})}\BibitemShut {NoStop}%
\bibitem [{\citenamefont {Skinner}\ \emph {et~al.}(2012)\citenamefont
  {Skinner}, \citenamefont {Chen},\ and\ \citenamefont
  {Shklovskii}}]{Skinner2012}%
  \BibitemOpen
  \bibfield  {author} {\bibinfo {author} {\bibfnamefont {Brian}\ \bibnamefont
  {Skinner}}, \bibinfo {author} {\bibfnamefont {Tianran}\ \bibnamefont {Chen}},
  \ and\ \bibinfo {author} {\bibfnamefont {B.~I.}\ \bibnamefont {Shklovskii}},\
  }\bibfield  {title} {\enquote {\bibinfo {title} {{Why is the bulk resistivity
  of topological insulators so small?}}}\ }\href@noop {} {\bibfield  {journal}
  {\bibinfo  {journal} {Phys. Rev. Lett.}\ }\textbf {\bibinfo {volume} {109}},\
  \bibinfo {pages} {176801} (\bibinfo {year} {2012})}\BibitemShut {NoStop}%
\bibitem [{\citenamefont {Tanaka}\ \emph {et~al.}(2013)\citenamefont {Tanaka},
  \citenamefont {Sato}, \citenamefont {Nakayama}, \citenamefont {Souma},
  \citenamefont {Takahashi}, \citenamefont {Ren}, \citenamefont {Novak},
  \citenamefont {Segawa},\ and\ \citenamefont {Ando}}]{Tanaka2013}%
  \BibitemOpen
  \bibfield  {author} {\bibinfo {author} {\bibfnamefont {Y.}~\bibnamefont
  {Tanaka}}, \bibinfo {author} {\bibfnamefont {T.}~\bibnamefont {Sato}},
  \bibinfo {author} {\bibfnamefont {K.}~\bibnamefont {Nakayama}}, \bibinfo
  {author} {\bibfnamefont {S.}~\bibnamefont {Souma}}, \bibinfo {author}
  {\bibfnamefont {T.}~\bibnamefont {Takahashi}}, \bibinfo {author}
  {\bibfnamefont {Zhi}\ \bibnamefont {Ren}}, \bibinfo {author} {\bibfnamefont
  {Mario}\ \bibnamefont {Novak}}, \bibinfo {author} {\bibfnamefont {Kouji}\
  \bibnamefont {Segawa}}, \ and\ \bibinfo {author} {\bibfnamefont {Yoichi}\
  \bibnamefont {Ando}},\ }\bibfield  {title} {\enquote {\bibinfo {title}
  {{Tunability of the $k$-space location of the Dirac cones in the topological
  crystalline insulator Pb$_{1-x}$Sn$_x$Te}},}\ }\href@noop {} {\bibfield
  {journal} {\bibinfo  {journal} {Phys. Rev. B}\ }\textbf {\bibinfo {volume}
  {87}},\ \bibinfo {pages} {155105} (\bibinfo {year} {2013})}\BibitemShut
  {NoStop}%
\bibitem [{\citenamefont {Dimmock}\ \emph {et~al.}(1966)\citenamefont
  {Dimmock}, \citenamefont {Melngailis},\ and\ \citenamefont
  {Strauss}}]{Dimmock1966}%
  \BibitemOpen
  \bibfield  {author} {\bibinfo {author} {\bibfnamefont {J.~O.}\ \bibnamefont
  {Dimmock}}, \bibinfo {author} {\bibfnamefont {Ivars}\ \bibnamefont
  {Melngailis}}, \ and\ \bibinfo {author} {\bibfnamefont {A.~J.}\ \bibnamefont
  {Strauss}},\ }\bibfield  {title} {\enquote {\bibinfo {title} {{Band structure
  and laser action in Pb$_x$Sn$_{1-x}$Te}},}\ }\href@noop {} {\bibfield
  {journal} {\bibinfo  {journal} {Phys. Rev. Lett.}\ }\textbf {\bibinfo
  {volume} {16}},\ \bibinfo {pages} {1193} (\bibinfo {year}
  {1966})}\BibitemShut {NoStop}%
\bibitem [{\citenamefont {Gao}\ and\ \citenamefont {Daw}(2008)}]{Gao2008}%
  \BibitemOpen
  \bibfield  {author} {\bibinfo {author} {\bibfnamefont {Xing}\ \bibnamefont
  {Gao}}\ and\ \bibinfo {author} {\bibfnamefont {Murray~S.}\ \bibnamefont
  {Daw}},\ }\bibfield  {title} {\enquote {\bibinfo {title} {{Investigation of
  band inversion in (Pb,Sn)Te alloys using ab initio calculations}},}\
  }\href@noop {} {\bibfield  {journal} {\bibinfo  {journal} {Phys. Rev. B}\
  }\textbf {\bibinfo {volume} {77}},\ \bibinfo {pages} {033103} (\bibinfo
  {year} {2008})}\BibitemShut {NoStop}%
\bibitem [{\citenamefont {Zhong}\ \emph {et~al.}(2014)\citenamefont {Zhong},
  \citenamefont {Schneeloch}, \citenamefont {Liu}, \citenamefont {Camino},
  \citenamefont {Tranquada},\ and\ \citenamefont {Gu}}]{Zhong2014}%
  \BibitemOpen
  \bibfield  {author} {\bibinfo {author} {\bibfnamefont {R.~D.}\ \bibnamefont
  {Zhong}}, \bibinfo {author} {\bibfnamefont {J.~A.}\ \bibnamefont
  {Schneeloch}}, \bibinfo {author} {\bibfnamefont {T.~S.}\ \bibnamefont {Liu}},
  \bibinfo {author} {\bibfnamefont {F.~E.}\ \bibnamefont {Camino}}, \bibinfo
  {author} {\bibfnamefont {J.~M.}\ \bibnamefont {Tranquada}}, \ and\ \bibinfo
  {author} {\bibfnamefont {G.~D.}\ \bibnamefont {Gu}},\ }\bibfield  {title}
  {\enquote {\bibinfo {title} {{Superconductivity induced by In substitution
  into the topological crystalline insulator
  ${\mathrm{Pb}}_{0.5}$${\mathrm{Sn}}_{0.5}\mathrm{Te}$}},}\ }\href@noop {}
  {\bibfield  {journal} {\bibinfo  {journal} {Phys. Rev. B}\ }\textbf {\bibinfo
  {volume} {90}},\ \bibinfo {pages} {020505} (\bibinfo {year}
  {2014})}\BibitemShut {NoStop}%
\bibitem [{\citenamefont {Shamshur}\ \emph {et~al.}(2010)\citenamefont
  {Shamshur}, \citenamefont {Parfen'ev}, \citenamefont {Chernyaev},\ and\
  \citenamefont {Nemov}}]{Shamshur2010}%
  \BibitemOpen
  \bibfield  {author} {\bibinfo {author} {\bibfnamefont {D.~V.}\ \bibnamefont
  {Shamshur}}, \bibinfo {author} {\bibfnamefont {R.~V.}\ \bibnamefont
  {Parfen'ev}}, \bibinfo {author} {\bibfnamefont {A.~V.}\ \bibnamefont
  {Chernyaev}}, \ and\ \bibinfo {author} {\bibfnamefont {S.~A.}\ \bibnamefont
  {Nemov}},\ }\bibfield  {title} {\enquote {\bibinfo {title} {{Low-temperature
  electrical conductivity and the superconductor-insulator transition induced
  by indium impurity states in (Pb$_{0. 5}$Sn$_{0. 5}$)$_{1-x}$In$_x$Te solid
  solutions}},}\ }\href@noop {} {\bibfield  {journal} {\bibinfo  {journal}
  {Phys. Solid State}\ }\textbf {\bibinfo {volume} {52}},\ \bibinfo {pages}
  {1815--1819} (\bibinfo {year} {2010})}\BibitemShut {NoStop}%
\bibitem [{\citenamefont {Ravich}\ and\ \citenamefont
  {Nemov}(2002)}]{Ravich2002}%
  \BibitemOpen
  \bibfield  {author} {\bibinfo {author} {\bibfnamefont {Yu.~I.}\ \bibnamefont
  {Ravich}}\ and\ \bibinfo {author} {\bibfnamefont {S.~A.}\ \bibnamefont
  {Nemov}},\ }\bibfield  {title} {\enquote {\bibinfo {title} {{Hopping
  conduction via strongly localized impurity states of indium in PbTe and its
  solid solutions}},}\ }\href@noop {} {\bibfield  {journal} {\bibinfo
  {journal} {Semiconductors}\ }\textbf {\bibinfo {volume} {36}},\ \bibinfo
  {pages} {1--20} (\bibinfo {year} {2002})}\BibitemShut {NoStop}%
\bibitem [{\citenamefont {Akimov}\ \emph {et~al.}(1983)\citenamefont {Akimov},
  \citenamefont {Brandt}, \citenamefont {Ryabova}, \citenamefont {Sokovishin},\
  and\ \citenamefont {Chudinov}}]{Akimov1983}%
  \BibitemOpen
  \bibfield  {author} {\bibinfo {author} {\bibfnamefont {B.~A.}\ \bibnamefont
  {Akimov}}, \bibinfo {author} {\bibfnamefont {N.~B.}\ \bibnamefont {Brandt}},
  \bibinfo {author} {\bibfnamefont {L.~I.}\ \bibnamefont {Ryabova}}, \bibinfo
  {author} {\bibfnamefont {V.~V.}\ \bibnamefont {Sokovishin}}, \ and\ \bibinfo
  {author} {\bibfnamefont {S.~M.}\ \bibnamefont {Chudinov}},\ }\bibfield
  {title} {\enquote {\bibinfo {title} {{Band edge motion in quantizing magnetic
  field and nonequilibrium states in Pb$_{1-x}$Sn$_x$Te alloys doped with
  In}},}\ }\href@noop {} {\bibfield  {journal} {\bibinfo  {journal} {J. Low
  Temp. Phys.}\ }\textbf {\bibinfo {volume} {51}},\ \bibinfo {pages} {9--32}
  (\bibinfo {year} {1983})}\BibitemShut {NoStop}%
\bibitem [{\citenamefont {Kim}\ \emph {et~al.}(2014)\citenamefont {Kim},
  \citenamefont {Xia},\ and\ \citenamefont {Fisk}}]{Kim2014}%
  \BibitemOpen
  \bibfield  {author} {\bibinfo {author} {\bibfnamefont {Dae-Jeong}\
  \bibnamefont {Kim}}, \bibinfo {author} {\bibfnamefont {J}~\bibnamefont
  {Xia}}, \ and\ \bibinfo {author} {\bibfnamefont {Z}~\bibnamefont {Fisk}},\
  }\bibfield  {title} {\enquote {\bibinfo {title} {{Topological surface state
  in the Kondo insulator samarium hexaboride}},}\ }\href@noop {} {\bibfield
  {journal} {\bibinfo  {journal} {Nat. Mater.}\ }\textbf {\bibinfo {volume}
  {13}},\ \bibinfo {pages} {466--470} (\bibinfo {year} {2014})}\BibitemShut
  {NoStop}%
\bibitem [{\citenamefont {Bansal}\ \emph {et~al.}(2012)\citenamefont {Bansal},
  \citenamefont {Kim}, \citenamefont {Brahlek}, \citenamefont {Edrey},\ and\
  \citenamefont {Oh}}]{Bansal2012}%
  \BibitemOpen
  \bibfield  {author} {\bibinfo {author} {\bibfnamefont {Namrata}\ \bibnamefont
  {Bansal}}, \bibinfo {author} {\bibfnamefont {Yong~Seung}\ \bibnamefont
  {Kim}}, \bibinfo {author} {\bibfnamefont {Matthew}\ \bibnamefont {Brahlek}},
  \bibinfo {author} {\bibfnamefont {Eliav}\ \bibnamefont {Edrey}}, \ and\
  \bibinfo {author} {\bibfnamefont {Seongshik}\ \bibnamefont {Oh}},\ }\bibfield
   {title} {\enquote {\bibinfo {title} {{Thickness-independent transport
  channels in topological insulator Bi$_2$Se$_3$ thin films}},}\ }\href@noop {}
  {\bibfield  {journal} {\bibinfo  {journal} {Phys. Rev. Lett.}\ }\textbf
  {\bibinfo {volume} {109}},\ \bibinfo {pages} {116804} (\bibinfo {year}
  {2012})}\BibitemShut {NoStop}%
\bibitem [{\citenamefont {Serbyn}\ and\ \citenamefont {Fu}(2014)}]{Serbyn2014}%
  \BibitemOpen
  \bibfield  {author} {\bibinfo {author} {\bibfnamefont {Maksym}\ \bibnamefont
  {Serbyn}}\ and\ \bibinfo {author} {\bibfnamefont {Liang}\ \bibnamefont
  {Fu}},\ }\bibfield  {title} {\enquote {\bibinfo {title} {{Symmetry breaking
  and Landau quantization in topological crystalline insulators}},}\
  }\href@noop {} {\bibfield  {journal} {\bibinfo  {journal} {Phys. Rev. B}\
  }\textbf {\bibinfo {volume} {90}},\ \bibinfo {pages} {035402} (\bibinfo
  {year} {2014})}\BibitemShut {NoStop}%
\bibitem [{\citenamefont {Hikami}\ \emph {et~al.}(1980)\citenamefont {Hikami},
  \citenamefont {Larkin},\ and\ \citenamefont {Nagaoka}}]{Hikami1980}%
  \BibitemOpen
  \bibfield  {author} {\bibinfo {author} {\bibfnamefont {Shinobu}\ \bibnamefont
  {Hikami}}, \bibinfo {author} {\bibfnamefont {Anatoly~I.}\ \bibnamefont
  {Larkin}}, \ and\ \bibinfo {author} {\bibfnamefont {Yosuke}\ \bibnamefont
  {Nagaoka}},\ }\bibfield  {title} {\enquote {\bibinfo {title} {{Spin-orbit
  interaction and magnetoresistance in the two dimensional random system}},}\
  }\href@noop {} {\bibfield  {journal} {\bibinfo  {journal} {Prog. Theor.
  Phys.}\ }\textbf {\bibinfo {volume} {63}},\ \bibinfo {pages} {707--710}
  (\bibinfo {year} {1980})}\BibitemShut {NoStop}%
\bibitem [{\citenamefont {Liu}\ \emph {et~al.}(2013)\citenamefont {Liu},
  \citenamefont {Duan},\ and\ \citenamefont {Fu}}]{Liu2014}%
  \BibitemOpen
  \bibfield  {author} {\bibinfo {author} {\bibfnamefont {Junwei}\ \bibnamefont
  {Liu}}, \bibinfo {author} {\bibfnamefont {Wenhui}\ \bibnamefont {Duan}}, \
  and\ \bibinfo {author} {\bibfnamefont {Liang}\ \bibnamefont {Fu}},\
  }\bibfield  {title} {\enquote {\bibinfo {title} {{Two types of surface states
  in topological crystalline insulators}},}\ }\href@noop {} {\bibfield
  {journal} {\bibinfo  {journal} {Phys. Rev. B}\ }\textbf {\bibinfo {volume}
  {88}},\ \bibinfo {pages} {241303} (\bibinfo {year} {2013})}\BibitemShut
  {NoStop}%
\bibitem [{\citenamefont {Assaf}\ \emph {et~al.}(2014)\citenamefont {Assaf},
  \citenamefont {Katmis}, \citenamefont {Wei}, \citenamefont {Satpati},
  \citenamefont {Zhang}, \citenamefont {Bennett}, \citenamefont {Harris},
  \citenamefont {Moodera},\ and\ \citenamefont {Heiman}}]{Assaf2014}%
  \BibitemOpen
  \bibfield  {author} {\bibinfo {author} {\bibfnamefont {Badih~A.}\
  \bibnamefont {Assaf}}, \bibinfo {author} {\bibfnamefont {Ferhat}\
  \bibnamefont {Katmis}}, \bibinfo {author} {\bibfnamefont {Peng}\ \bibnamefont
  {Wei}}, \bibinfo {author} {\bibfnamefont {Biswarup}\ \bibnamefont {Satpati}},
  \bibinfo {author} {\bibfnamefont {Zhan}\ \bibnamefont {Zhang}}, \bibinfo
  {author} {\bibfnamefont {Steven~P.}\ \bibnamefont {Bennett}}, \bibinfo
  {author} {\bibfnamefont {Vincent~G.}\ \bibnamefont {Harris}}, \bibinfo
  {author} {\bibfnamefont {Jagadeesh~S.}\ \bibnamefont {Moodera}}, \ and\
  \bibinfo {author} {\bibfnamefont {Don}\ \bibnamefont {Heiman}},\ }\bibfield
  {title} {\enquote {\bibinfo {title} {{Quantum coherent transport in SnTe
  topological crystalline insulator thin films}},}\ }\href@noop {} {\bibfield
  {journal} {\bibinfo  {journal} {Appl. Phys. Lett.}\ }\textbf {\bibinfo
  {volume} {105}},\ \bibinfo {pages} {102108} (\bibinfo {year}
  {2014})}\BibitemShut {NoStop}%
\bibitem [{\citenamefont {Pletikosi{\'c}}\ \emph {et~al.}(2014)\citenamefont
  {Pletikosi{\'c}}, \citenamefont {Gu},\ and\ \citenamefont
  {Valla}}]{Pletikosic2014}%
  \BibitemOpen
  \bibfield  {author} {\bibinfo {author} {\bibfnamefont {Ivo}\ \bibnamefont
  {Pletikosi{\'c}}}, \bibinfo {author} {\bibfnamefont {Genda}\ \bibnamefont
  {Gu}}, \ and\ \bibinfo {author} {\bibfnamefont {Tonica}\ \bibnamefont
  {Valla}},\ }\bibfield  {title} {\enquote {\bibinfo {title} {{Inducing a
  Lifshitz Transition by Extrinsic Doping of Surface Bands in the Topological
  Crystalline Insulator Pb$_{1-x}$Sn$_x$Se}},}\ }\href@noop {} {\bibfield
  {journal} {\bibinfo  {journal} {Phys. Rev. Lett.}\ }\textbf {\bibinfo
  {volume} {112}},\ \bibinfo {pages} {146403} (\bibinfo {year}
  {2014})}\BibitemShut {NoStop}%
\bibitem [{\citenamefont {Zhong}\ \emph {et~al.}(2013)\citenamefont {Zhong},
  \citenamefont {Schneeloch}, \citenamefont {Shi}, \citenamefont {Xu},
  \citenamefont {Zhang}, \citenamefont {Tranquada}, \citenamefont {Li},\ and\
  \citenamefont {Gu}}]{Zhong2013}%
  \BibitemOpen
  \bibfield  {author} {\bibinfo {author} {\bibfnamefont {R.~D.}\ \bibnamefont
  {Zhong}}, \bibinfo {author} {\bibfnamefont {J.~A.}\ \bibnamefont
  {Schneeloch}}, \bibinfo {author} {\bibfnamefont {X.~Y.}\ \bibnamefont {Shi}},
  \bibinfo {author} {\bibfnamefont {Z.~J.}\ \bibnamefont {Xu}}, \bibinfo
  {author} {\bibfnamefont {C.}~\bibnamefont {Zhang}}, \bibinfo {author}
  {\bibfnamefont {J.~M.}\ \bibnamefont {Tranquada}}, \bibinfo {author}
  {\bibfnamefont {Q.}~\bibnamefont {Li}}, \ and\ \bibinfo {author}
  {\bibfnamefont {G.~D.}\ \bibnamefont {Gu}},\ }\bibfield  {title} {\enquote
  {\bibinfo {title} {{Optimizing the superconducting transition temperature and
  upper critical field of Sn$_{1-x}$In$_{x}$Te}},}\ }\href@noop {} {\bibfield
  {journal} {\bibinfo  {journal} {Phys. Rev. B}\ }\textbf {\bibinfo {volume}
  {88}},\ \bibinfo {pages} {020505} (\bibinfo {year} {2013})}\BibitemShut
  {NoStop}%
\end{thebibliography}

%

\end{document}